\documentclass[10pt]{wlscirep}
\usepackage[utf8]{inputenc}
\usepackage[T1]{fontenc}
\usepackage{amsmath}
\usepackage{hyperref}
\usepackage{lineno}
\usepackage{float}
\usepackage{orcidlink}
\DeclareMathSymbol{,}{\mathpunct}{operators}{"2C}

\title{Precision measurement of the return distribution property of the Chinese stock market index}

\author[1,*,+]{Peng Liu\orcidlink{0000-0002-9115-2055}}
\author[2,+]{Yanyan Zheng\orcidlink{0000-0002-1092-072X}}
\affil[1]{School of Information, Xi'an University of Finance and Economics, Xi'an 710100, Shaanxi, P.R. China}
\affil[2]{School of Management, Xi'an Polytechnic University, Xi'an 710048, Shaanxi, P.R. China}

\affil[*]{corresponding author: \href{mailto:pengliu@xaufe.edu.cn}{pengliu@xaufe.edu.cn}}

\affil[+]{these authors contributed equally to this work}

\begin{abstract}
In econophysics, the analysis of the return distribution of a financial asset using statistical physics methods is a long-standing and important issue.
This paper systematically conducts an analysis of composite index 1 min datasets over a 17-year period (2005-2021) for both the Shanghai and Shenzhen stock exchanges.
To reveal the differences between  Chinese and mature stock markets, we precisely measure the property of the return distribution of the composite index over the time scale $\Delta t$, which ranges from 1 min to almost 4,000 min.
The main findings are as follows:
(1) The return distribution presents a leptokurtic, fat-tailed, and almost symmetrical shape that is similar to that of mature markets.
(2) The central part of the return distribution is described by the symmetrical L\'{e}vy $\alpha$-stable process,
with a stability parameter comparable with a value of about 1.4, which was extracted for the U.S. stock market.
(3) The return distribution can be described well by  Student's t-distribution within a wider return range than the L\'{e}vy $\alpha$-stable distribution.
(4) Distinctively, the stability parameter shows a potential change when $\Delta t$ increases,
and thus a crossover region at 15 $< \Delta t <$ 60 min is observed.
This is different from the finding in the U.S. stock market that a single value of about 1.4 holds over 1 $\le \Delta t \le$ 1,000 min.
(5) The tail distribution of returns at small $\Delta t$ decays as an asymptotic power-law with an exponent of about 3, which is a widely observed value in mature markets.
However, it decays exponentially when $\Delta t \ge$ 240 min, which is not observed in mature markets.
(6) Return distributions gradually converge to a normal distribution as $\Delta t$ increases. This observation is different from the finding of a critical $\Delta t =$ 4 days in the U.S. stock market.
\\
\\
\textbf{Keywords:} econophysics; sociophysics; return distribution; Chinese stock market index; power-law
\end{abstract}

\begin{document}

\flushbottom
\maketitle

\thispagestyle{empty}
{\noindent}\rule[0pt]{17.5cm}{0.05em}

\section{Introduction}
Econophysics is an emerging interdisciplinary field.
It investigates economic and financial problems through the models, methods, and concepts adopted in physics, especially statistical physics{\cite{NP-1969, Kutner-2019, Ribeiro-2020, EPL, entropy}}.
Among the most important and remarkably interesting studies in the econophysics field, the price fluctuation of assets in the financial market
has been intensively investigated in both empirical and theoretical ways since 1900
{\cite{Stanley-book-2000, Bouchaud-2000, Malevergne-2006, Bachelier-1990, Fama-1970, Mandelbrot-1960, Mandelbrot-1963-1, Mandelbrot-1963-2, Mandelbrot-1967, Akgiray-1988, Multifractal, Merton-1976, Variance-gamma, Kou-2002, CGMY, Schoutens-2001, Heston-1993, Dupire-1994, Chourdakis-2005, Gatheral-2015}}.

The stock market is a complex financial system in which the traders, assets, and many unforeseen external factors interact with each other non-linearly; thus, it is extremely hard to write down a dynamical equation among these elements.
Fortunately, the price fluctuations of individual stocks and market indices provide us with a powerful tool to understand its dynamics{\cite{Kutner-2019, Stanley-book-2000}}.
Fluctuations are often quantified by a logarithmic return over a time scale of $\Delta t$ that can be mathematically defined as follows:
\begin{equation}
\label{return}
R_{\Delta t} \left(t\right) = \ln \frac{S\left(t\right)}{S\left(t - \Delta t\right)}
\end{equation}
where $S\left(t \right)$ denotes the time series of a company stock price or market index.
The return is of a great key role in asset pricing and is at the core of financial risks through multifractal analysis{\cite{Stanley-book-2000, Bouchaud-2000, Malevergne-2006, Multifractal}}.

In the late 20th century, huge amounts of data from the stock market are available for scholars due to the rapid development of computer technology{\cite{Stanley-book-2000, Kutner-2019}}.
This development allows physicists to analyze precisely the properties of the return of a financial asset using the methodology developed for statistical physics.
In 1995, a paper published in Nature analyzed the return distribution of the Standard \& Poor's 500 (S\&P 500) index over the 6-year period (1984-1989){\cite{Nature-1995}}.
This work found that the central region of the return distribution
can be well described by a truncated L\'{e}vy  stable symmetrical distribution{\cite{PRL-1994}} with an index of $\alpha \approx$ 1.4
(comparable with $\alpha \approx$ 1.5 for the income distribution{\cite{Ribeiro-2020}} and $\alpha \approx$ 1.7 for the distribution of the fluctuation of cotton price{\cite{Mandelbrot-1963-2}}).
More importantly, this study observed the scaling behavior of the probability density over three orders of magnitude of $\Delta t$.
Four years later, 
the same team conducted more detailed studies on the stock indices{\cite{PRE-1999-1}} and individual company stocks{\cite{EPJB-1998, PRE-1999-2}} in the U.S. market.
These new works found a universal asymptotic inverse cubic power-law in the return distribution tails for both the S\&P 500 index and individual company stocks (the power-law has been observed in many natural and social complex systems{\cite{Stretched-exp}}).
They also observed a critical point of $\Delta t$ below which the tail distributions retain a similar power-law, and gradually converge to Gaussian distribution otherwise
 ( $\Delta t \approx$ 4 days and $\Delta t \approx$ 16 days for market index and individual company stocks, respectively) {\cite{PRE-1999-1, PRE-1999-2, Nature-2003}}.
Further studies illustrated similar scaling behavior in other mature stock markets,
such as the market in England, France, Germany, Mexico, and Japan{\cite{PRE-1999-1, PRE-1999-2, PRE-2008, Stanley-2008-review, Germany-1996, Mexican-2005, Mexican-2012}}.

Although mature stock markets seem to show a universality of power-law scaling behavior,
many exceptions exist in other stock markets.
Studies on the stocks traded in the Australian Stock Exchange and the daily WIG index of the Warsaw Stock Exchange
showed that tail distributions follow power-law, with the exponent being significantly different from 3{\cite{Australia-2002, Australia-2004, poland-2001}}. 
As for the Indian stock market, a study on the daily returns of the 49 largest stocks indicated{\cite{India-2004}} that the tail distributions decay exponentially as $e^{-\beta r}$.
However, new studies in 2007 and 2008{\cite{India-2007, India-2008}} found
that the distributions of fluctuations of the individual stock prices, the Nifty index, and the Sensex index follow the asymptotic power-law with exponent $\alpha \approx$ 3.
For the Hong Kong stock market, research has also been conducted and delivered different results from research based on the U.S. stock market{\cite{HK-2000, HangSeng-2001}}.
These non-unified results make it difficult to understand the dynamical property of the stock market.
It seems to be dependent on the degree of development of a specific financial market{\cite{diff-2003}}.

The Chinese Mainland stock market, the largest emerging financial market in the world,
has different trading rules and government regulations compared with other developed financial markets,
such as T + 1 trading, intraday price limits, and IPO policy{\cite{price-limit-2015}}.
These differences may result in different interactions among traders, assets, and external factors in the Chinese stock market.
Thus, it is of great importance to understand the dynamical properties of the Chinese stock market via return distributions.
However, the previous analyses on the return distribution for the Chinese stock market were conducted about 10 years ago
and did not obtain conclusive results because of the limitation of data statistics.
In 2005, scholars analyzed the data of 104 individual stocks listed on the Shanghai Stock Exchange (SSE) and Shenzhen Stock Exchange (SZSE)
and found the tail distributions of daily stock price returns follow the power-law, with the exponent being significantly different from that in the U.S. stock market{\cite{China-2005}}.
They also stated that the distributions of returns are asymmetrical, but almost symmetrical distributions were observed in the U.S. stock market{\cite{China-2005}}.
A similar analysis of the SSE Composite Index (SSECI) and the SZSE Component Index in 2007 observed
a symmetrical return distribution with a power-law exponent of less than 3 over 1 $\le \Delta t \le$ 60 min{\cite{China-2007}}.
In contrast, a more detailed analysis of the 1 min data and 1-day data of the SSECI in 2008 showed{\cite{China-2008}} that the power-law exponent is systematically larger than 3.
Subsequently, a study on the tick-by-tick data from 23 individual stocks listed in SZSE argued that return distribution can be well fitted with the Student's t-distribution{\cite{China-2008-1, China-2010}}, 
which is different from the truncated L\'{e}vy stable process model{\cite{PRL-1994, Nature-1995}}.
In 2010, a study of the SSE 50 index and SZSE 100 index revealed{\cite{China-2010-1}}
that tail distributions obey the power-law when $\Delta t <$ 1 week and follow exponential decay otherwise.

In the past 20 years, more high-frequency data on the Chinese stock market have been accumulated for analysis.
These data provide us with a good opportunity to precisely measure the properties of return distributions.
To shed light on the understanding of the dynamical property of the Chinese stock market and help clarify the confusion stated above,
this paper analyzes 1 min datasets recorded for the SSECI and the SZSE Composite Index (SZSECI) over the 17-year period (January 4, 2005 to December 31, 2021),
using the methods and concepts adopted in statistical physics.

\section{Datasets}
\begin{figure}[H]
\centering
\includegraphics[width=0.625\linewidth]{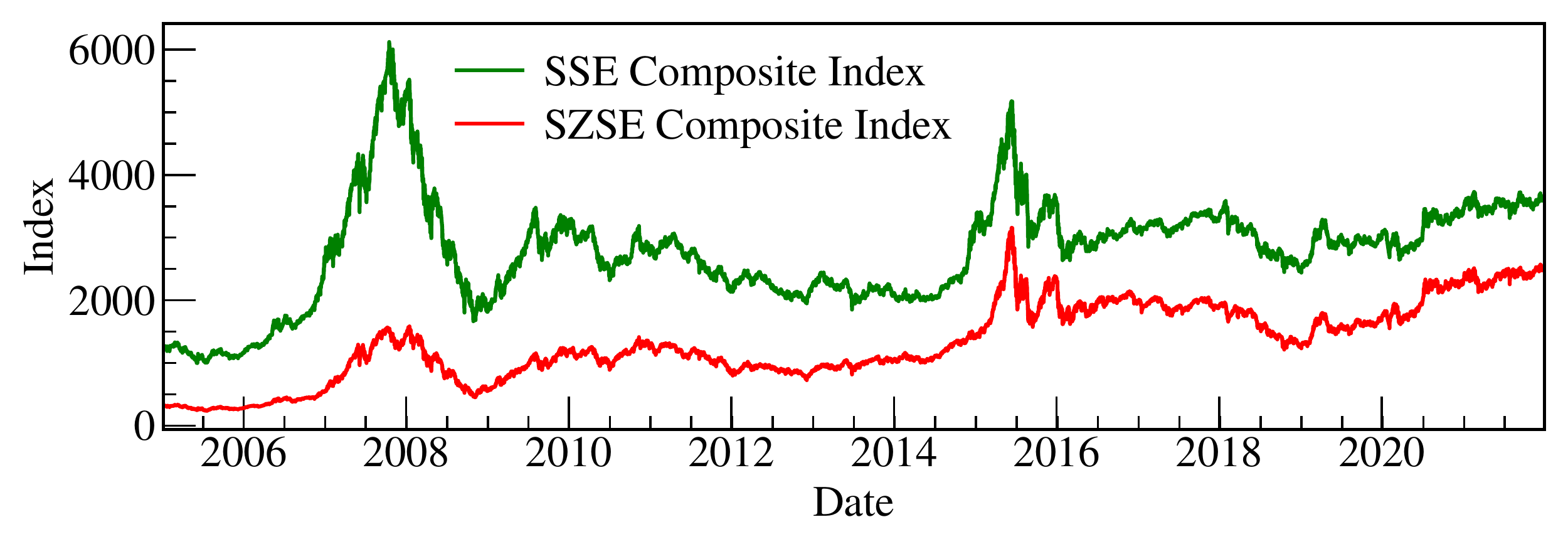}
\caption{\textbf{The time series of the SSECI and the SZSECI.} The 1 min datasets over the 17-year period (January 4, 2005 to December 31, 2021) analyzed in this paper are shown. The higher one and lower one represent the SSECI and the SZSECI, respectively.}
\label{fig0}
\end{figure}
\newpage
This paper analyzes 1 min datasets over the 17-year period (January 4, 2005 to December 31, 2021)
for both the SSECI and the SZSECI.
The SSE and SZSE, the only two stock exchanges in Mainland China, were established in late 1990.
The SSECI comprises all stocks of A-shares and B-shares listed and traded on SSE.
Similarly, the SZSECI consists of all stocks listed and traded on SZSE.
Both indices aim to reflect the overall Chinese stock market performance and are calculated by the capitalization-weighted method.
Both exchanges trade during 9:30-11:30 and 13:00-15:00 of a trading day and are closed on the weekend and national holidays.
When we construct a time series $S\left(t \right)$,
we first skip the non-trading days and the period of 11:30-13:00 on trading days,
and then connect 9:00 to 15:00 of the previous trading day.
The time series of indices with other time scales are constructed using the 1 min datasets (991,680 records for each index).

\section{Results and discussion}
\begin{figure}[H]
\centering
\includegraphics[width=1.0\linewidth]{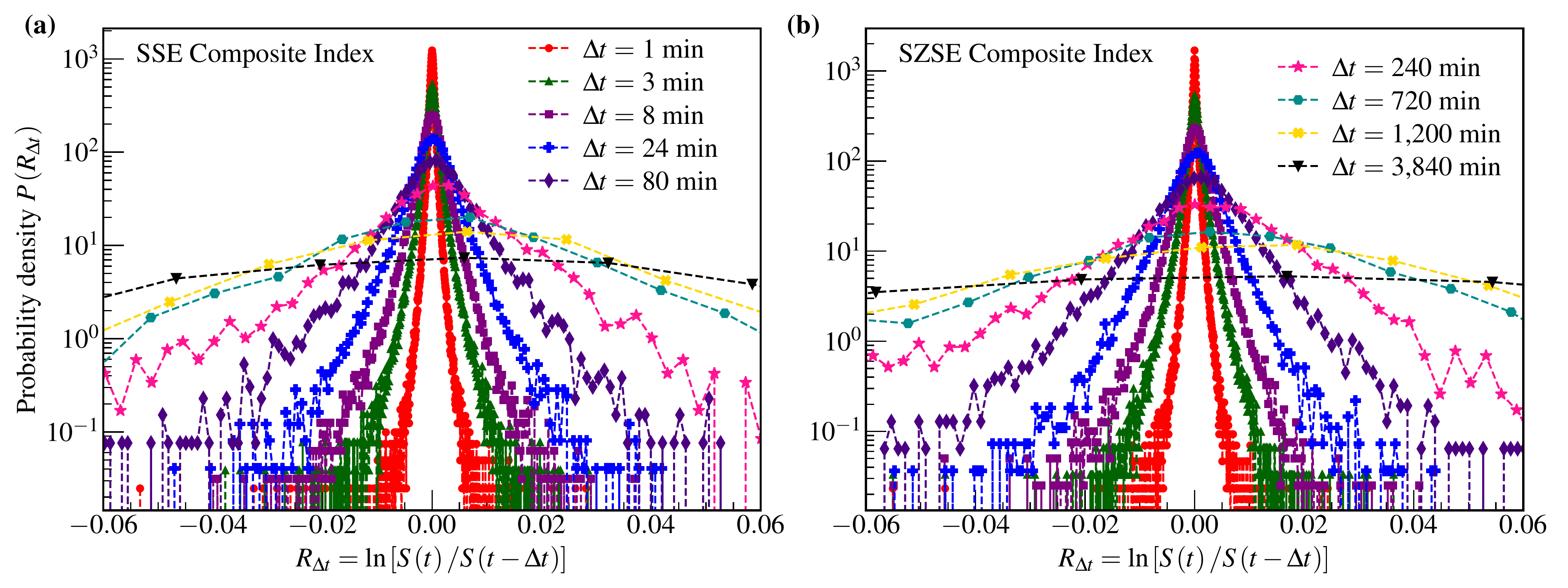}
\caption{\textbf{Probability density of returns over time intervals ranging from 1 min to 3,840 min.}
Return $R_{\Delta t}$ is defined as $\ln \left[S\left( t \right) / S\left(t - \Delta t \right)\right]$, where $S(t)$ refers to the time series of the SSECI (panel a) or the SZSECI (panel b).
Different markers represent data of returns with different time intervals $\Delta t$.
These two panels share a common legend. 
It is evident that these distributions expand as $\Delta t$ increases.}
\label{fig1}
\end{figure}

To provide an overview of the statistical property of returns,
Fig. \ref{fig1} shows the probability density functions (PDFs) of returns of over 1 $\le \Delta t \le$ 3,840 min for both the SSECI and the SZSECI.
It is evident from Fig. \ref{fig1} that the PDFs of both indices have a similar shape.
To study the shape quantitatively, we examine the skewness and kurtosis and the corresponding statistical significance tests{\cite{skewtest, kurtosistest}}.
These examinations show that our data present slightly negative skewness with statistical significance, but its most central parts are symmetrical.
These examinations also show that Fisher's kurtosis is larger than 0, with very high statistical significance.
Additionally, as can be seen in this figure, these distributions are leptokurtic, fat-tailed, and almost symmetrical.

Previous works have illustrated that the central parts of the return PDFs shown in Fig. \ref{fig1}
can be well described by the L\'{e}vy $\alpha$-stable process{\cite{Stanley-book-2000, PRL-1994, Nature-1995, China-2007, China-2008}}.
The symmetrical L\'{e}vy $\alpha$-stable PDF is mathematically written as
\begin{equation}
P_{\alpha} \left(R_{\Delta t}, \Delta t\right) = \frac{1}{\pi}\int_{0}^{+\infty}e^{-\gamma \Delta t \left|q\right|^{\alpha}} \cos (qR_{\Delta t})dq
\label{Levydis}
\end{equation}
where $P_{\alpha}$ refers to PDF, $R_{\Delta t}$ is the return defined by Eq. (\ref{return}),
$\Delta t$ denotes time scale (in Eqs. (\ref{Levydis}) - (\ref{transformation2}), to make $\Delta t$ dimensionless,  we let $\Delta t$ equal the time scale divided by 1 minute),
$\alpha$ ($0 < \alpha \le 2$, stability parameter, also known as index) is a key parameter for L\'{e}vy $\alpha$-stable distribution,
and $\gamma$ is the scale parameter.
In Eq. (\ref{Levydis}), $e^{-\gamma \Delta t \left|q\right|^{\alpha}}$ is the characteristic function.
According to Eq. (\ref{Levydis}), the PDF of $R_{\Delta t} = $ 0 is
\begin{equation}
P_{\alpha}(R_{\Delta t} = 0, \Delta t) = \frac{\Gamma\left(1/\alpha\right)}{\pi \alpha\left(\gamma \Delta t\right)^{1/\alpha}}
\label{P0}
\end{equation}
where $\Gamma$ denotes the Gamma function.

\begin{figure}[H]
\centering
\includegraphics[width=1.0\linewidth]{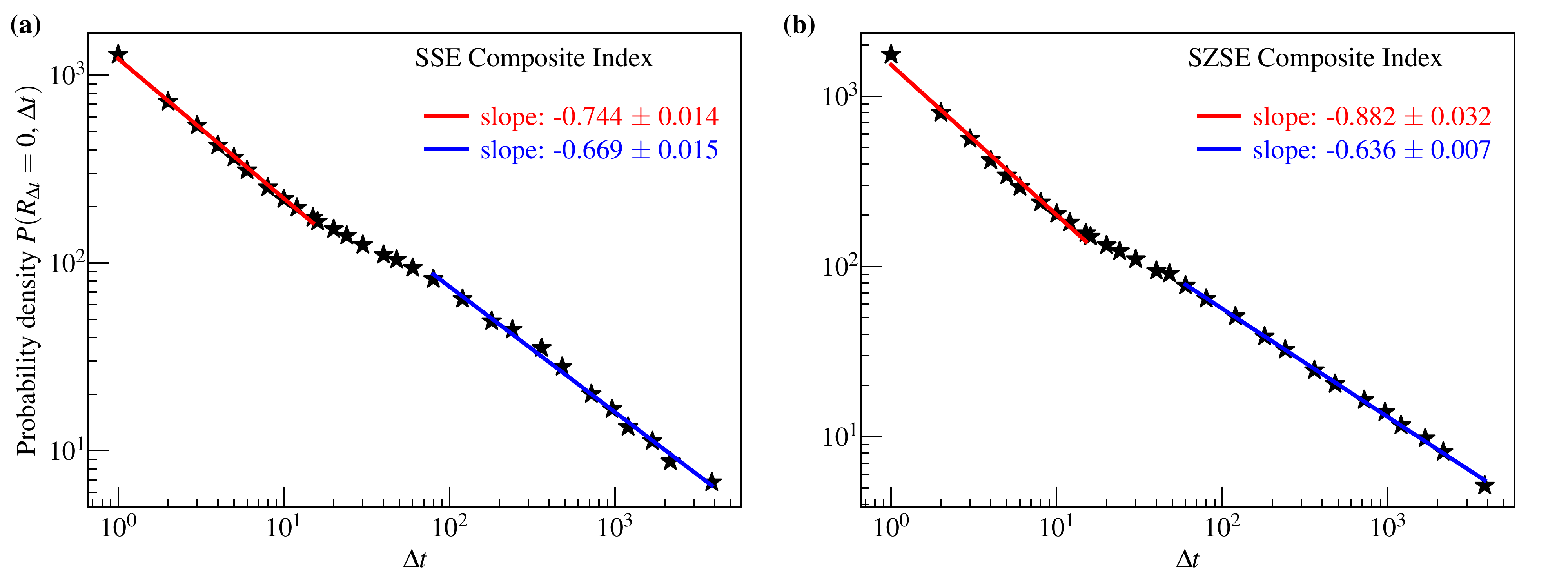}
\caption{\textbf{Probability density of the return $R_{\Delta t} = 0$ as a function of the time interval $\Delta t$.}
The black stars are the data points.
The red and blue lines are straight-line fits to the data points over 1 $\le \Delta t \le$ 15 and 60 $\le \Delta t \le$ 3,840, respectively.
The fit results of slopes for red and blue lines are also shown. Similar scaling behavior is observed for both the SSECI (panel a) and the SZSECI (panel b).
From this figure, the key parameter $\alpha$ characterizing the L\'{e}vy $\alpha$-stable process is extracted (see main text for details). A potential crossover region at 15 $< \Delta t <$ 60 is observed here.}
\label{fig2}
\end{figure}

Next, we use the approach proposed by Ref. {\cite{Nature-1995}} to extract the stability parameter $\alpha$ from the data shown in Fig. \ref{fig1}.
According to Eq. (\ref{P0}), the parameter $\alpha$ equals the negative of the reciprocal of the slope shown in Fig. \ref{fig2}.
The $\alpha$ values are as follows: 1.34 $\pm$ 0.03 (SSECI) and 1.13 $\pm$ 0.04 (SZSECI) over 1 $\le \Delta t \le$ 15 min,
and 1.49 $\pm$ 0.03 (SSECI) and 1.57 $\pm$ 0.02 (SZSECI) over 60 $\le \Delta t \le$ 3,840 min.
Such values of $\alpha$ extracted from our data are consistent with 0 $< \alpha \le$ 2 and comparable with 1.40 $\pm$ 0.05 which is extracted from the U.S. stock market{\cite{Nature-1995}}.
A potential crossover region at 15 $< \Delta t <$ 60 min for these two indices is observed.
This potential crossover region is not observed in the U.S. stock market in which a single fitting holds over 1 $\le \Delta t \le$ 1,000 min{\cite{Nature-1995}},
and also in the previous similar studies regarding the Chinese Stock market{\cite{China-2007, China-2008}}.
We skip the first few minutes to an hour for each trading day to see whether this crossover region disappears,
but it exists.
The overnight return also cannot contribute to this phenomenon since the data points with $\Delta t \le$ 60 min do not include the overnight effect.
By removing the data affected by extreme events,
such as the global financial crisis of 2007-2008, the 2015-2016 Chinese stock market turbulence, and the COVID-19 global pandemic{\cite{COVID-19, COVID-19-2}}, 
we observe that those extreme events have no contribution to this crossover region.
Therefore, this potential crossover region may indicate an underlying dynamical behavior of the Chinese stock market that differs from the U.S. stock market.

The symmetrical L\'{e}vy $\alpha$-stable distribution shown in Eq. (\ref{Levydis}) will collapse on the $\Delta t = $1 distribution under the transformations below.

\begin{equation}
	R_{s} = R_{\Delta t} \left(\Delta t\right)^{-\frac{1}{\alpha}}
	\label{transformation1}
\end{equation}

\begin{equation}
	P_{\alpha}\left(R_{s}, 1\right) = P_{\alpha}\left(R_{\Delta t}, \Delta t\right)  \left(\Delta t\right)^{\frac{1}{\alpha}}
	\label{transformation2}
\end{equation}
where $R_{s}$ denotes rescaled return, and $R_{\Delta t}$ and $P_{\alpha}\left(R_{\Delta t}, \Delta t\right)$ are defined by Eqs. (\ref{return}) and (\ref{Levydis}), respectively.

\begin{figure}[H]
\centering
\includegraphics[width=1.0\linewidth]{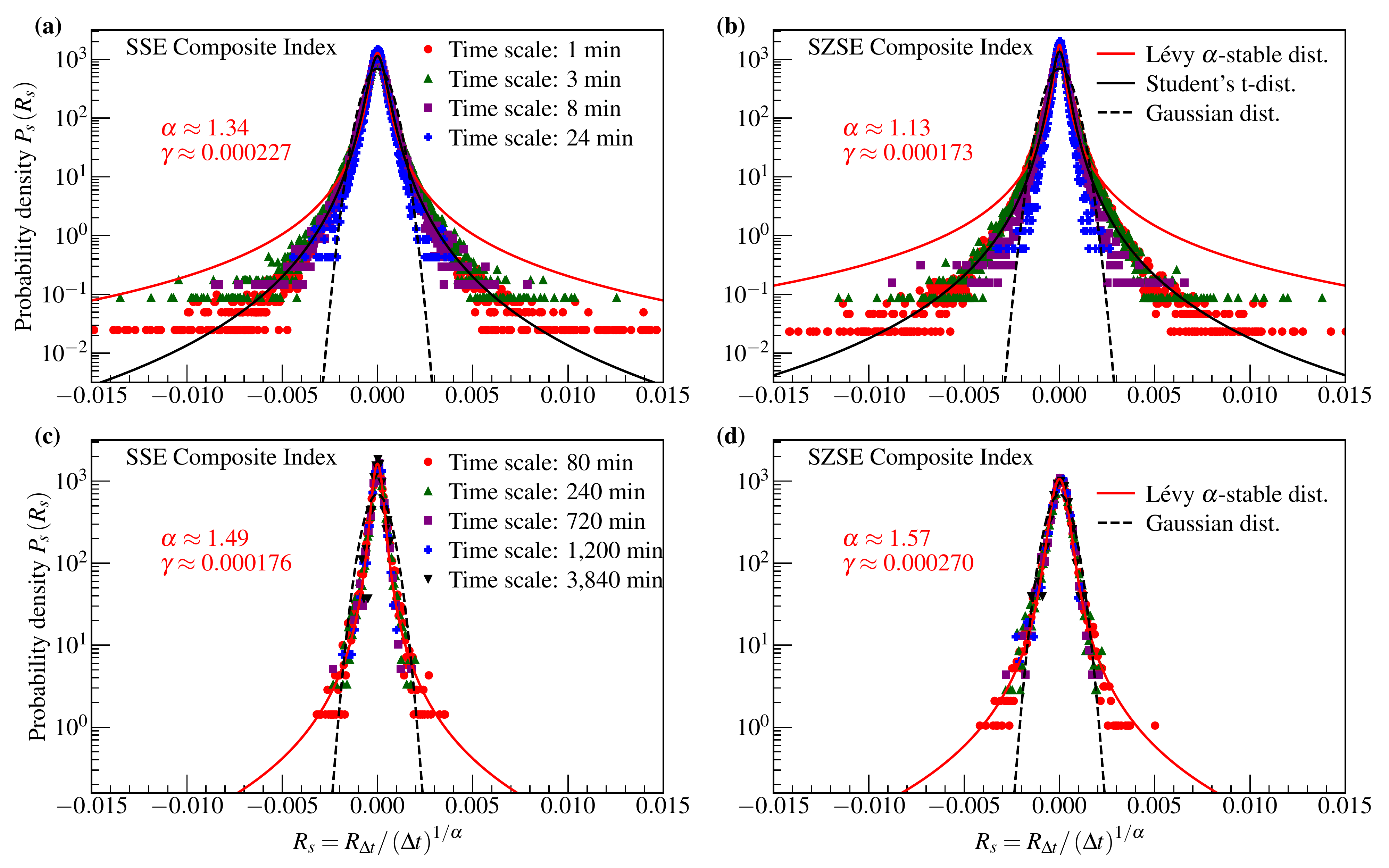}
\caption{\textbf{The comparison of rescaled PDFs of returns $R_{s}$ with theoretical models.}
The colored markers are data points with different time scales.
The dashed black curves are Gaussian distributions with a mean of 0 and a standard deviation of 1 min data.
The solid black curves are Student's t-distribution fits to 1 min data. 
The solid red curves are symmetrical L\'{e}vy $\alpha$-stable distributions with the observed parameters, as shown by the red text (see main text for details). 
The top two panels and the bottom two panels share a common legend, respectively.}
\label{fig3}
\end{figure}

Fig. \ref{fig3} shows the rescaled PDFs of returns $R_{s}$ with the stability parameter $\alpha$ extracted from Fig. \ref{fig2}.
An obvious collapse of distributions with a large $\Delta t$ is observed here.
However, only central parts of distributions with a larger $\Delta t$ overlap with the $\Delta t = $1 min data.
The red curves are the symmetrical L\'{e}vy $\alpha$-stable distributions with the parameters $\alpha$ obtained in Fig. \ref{fig2}.
Note that these red curves are not simple fits to data.
Their scale factors $\gamma$ are obtained using Eq. (\ref{P0}) and
the experimental $P\left(0\right)$ for $\Delta t \le$ 15 min 
or the extrapolation of $P\left(0\right)$ using the straight-line fits for $\Delta t \ge$ 60 min in Fig. \ref{fig2}.
The symmetrical L\'{e}vy $\alpha$-stable distributions with the parameters extracted from Fig. \ref{fig2} show good agreement with data in the central parts.
From the two aspects discussed above, we could conclude that the symmetrical L\'{e}vy $\alpha$-stable process describes a part of the dynamical properties of the Chinese stock market.
In Fig. \ref{fig3}, the tail distributions of data are larger than the Gaussian distribution and smaller than the L\'{e}vy $\alpha$-stable distribution.
Thus, we tried using the Student's t-distribution to fit data, as demonstrated by the solid black curves shown in panels a and b which have sufficient data.
The fit results show that the Student's t-distribution can describe the data at a wider range than the L\'{e}vy $\alpha$-stable distribution well, which could be explained by the so-called non-extensive statistical framework{\cite{nonextensive-1, nonextensive-2}}.

\newpage

Given the discussion above, it is essential to obtain a detailed study of the tail distribution.
To compare tails with different $\Delta t$, we introduce the normalized return $r_{\Delta t}$
\begin{equation}
r_{\Delta t} = \frac{R_{\Delta t} - \left<R_{\Delta t}\right>_{T}}{V}
\label{normalizedR}
\end{equation}
where $\left<R_{\Delta t}\right>_{T}$ is the average of returns $R_{\Delta t}$ over the entire time $T$, and $V$ is the volatility of the $R_{\Delta t}$ time series.

\begin{figure}[H]
\centering
\includegraphics[width=1.0\linewidth]{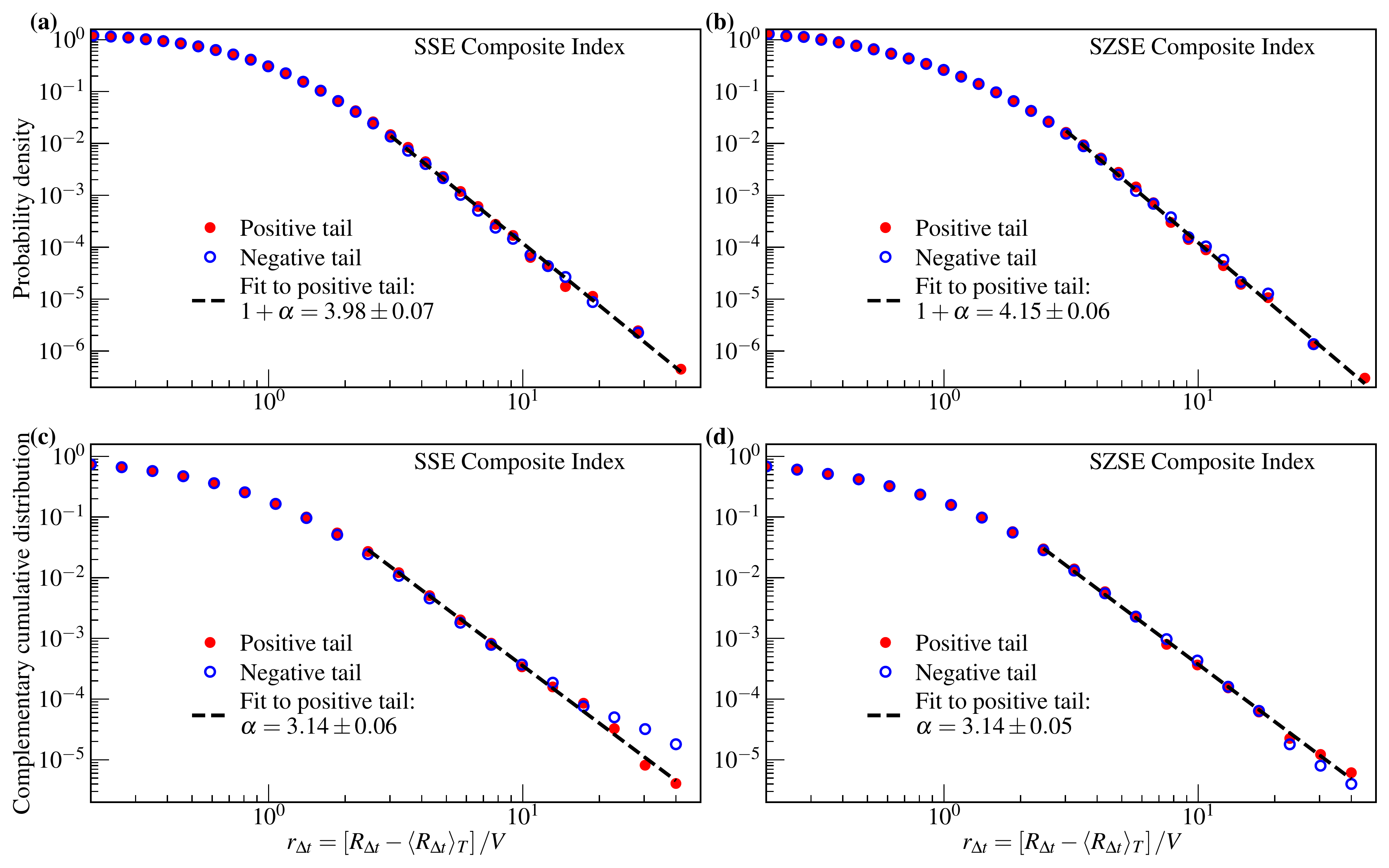}
\caption{\textbf{PDF and CCDF of the normalized return tails with a time scale of 1 min.}
Panels a and c reflect the SSECI, and panels b and d represent the SZSECI.
The red full circles and blue empty circles represent positive and negative tails, respectively.
Dashed black straight lines are straight-line fits to data.
Fit results of power-law exponents $\alpha$ are shown.
It is evident that the positive and negative tails show very similar behavior and follow a similar asymptotic power-law decay.}
\label{fig4}
\end{figure}

Fig. \ref{fig4} shows the PDF and the complementary cumulative distribution function (CCDF) of the $\Delta t$ = 1 min tail of normalized returns $r_{\Delta t}$ in a log--log style.
The PDF of the tail follows a power-law decay in the form of $r_{\Delta t}^{-(1 + \alpha)}$,
as shown in panels a and b.
Naturally, the CCDF follows the form of $r^{-\alpha}_{\Delta t}$,
as shown in panels c and d.
Similar behavior for both positive and negative tails is observed obviously.
We use a straight-line fit to extract the exponent $\alpha$,
as shown by the dashed black lines.
The exponents extracted from PDF are consistent with those from CCDF within acceptable errors.
For the SSECI positive tail,
the average (weighted by the reciprocal of squared errors) of exponents $\alpha$ extracted from fits to PDF and CCDF is 3.07 $\pm$ 0.05.
For the SZSECI positive tail, that value is 3.14 $\pm$ 0.04.
The values of the SSECI and the SZSECI are in agreement with each other within errors, and well as outside the L\'{e}vy $\alpha$-stable process.

\begin{figure}[H]
\centering
\includegraphics[width=1.0\linewidth]{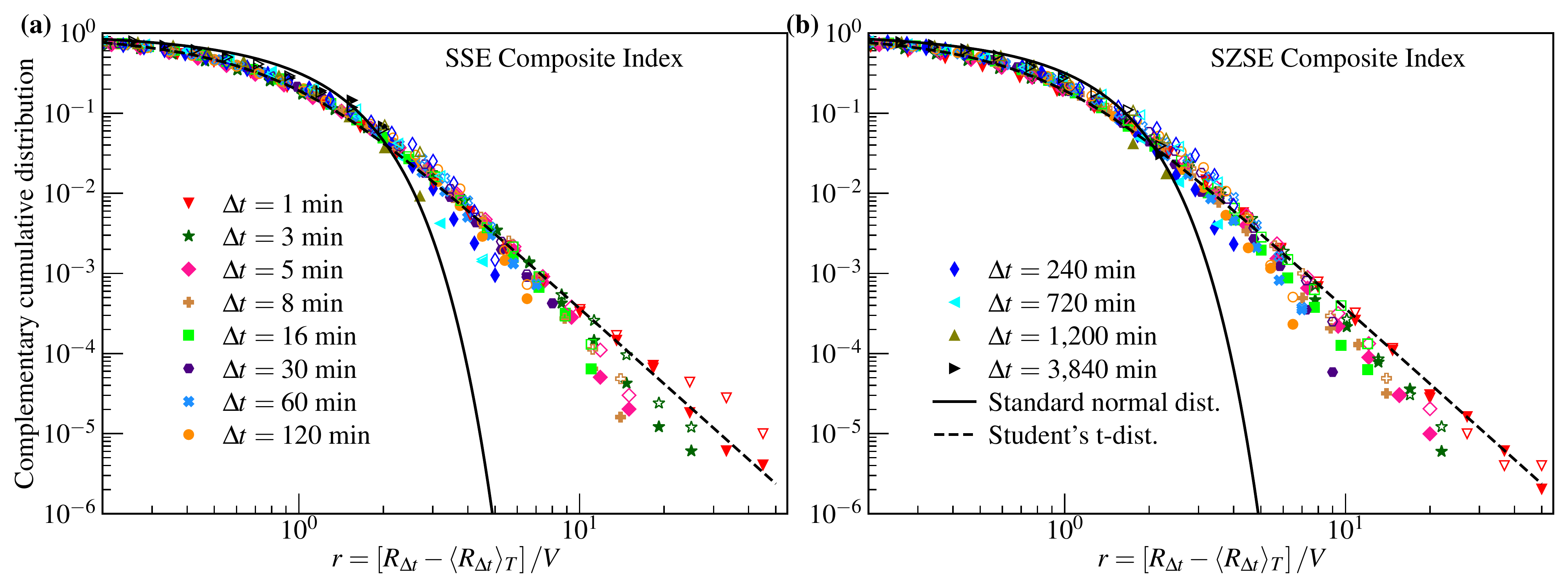}
\caption{\textbf{CCDF of the normalized return tails with different time scales $\Delta t$ in log--log plot.}
Panels a and b represent the SSECI and the SZSECI, respectively.
These two panels share a common legend.
The colored solid markers denote positive tails,
and the corresponding open markers represent negative tails.
The solid and dashed black curves represent the standard normal distribution and the Student's t-distribution (with a degree of freedom of 3.14 and a standard deviation of 1), respectively.
Scaling behavior in tail distributions is observed here, and the tails with small $\Delta t$ follow an asymptotic power-law decay.}
\label{fig5}
\end{figure}

Fig. \ref{fig5} compares the CCDF of the normalized returns with 1 $\le \Delta t \le$ 3,840 min for positive and negative tails.
Here, the theoretical values of standard normal distribution and Student's t-distribution are also drawn for a comparison.
It can be seen from Fig. \ref{fig5} that these tails with small time scales follow an asymptotic power-law decay, but the tails gradually deviate from the power-law when $ \Delta t$ becomes longer.
For the case of short $ \Delta t$,
the tail distributions are fitted by the power-law and the Student's t-distribution; thus, the exponents  $\alpha$ are extracted.
The exponents $\alpha$ extracted from these two functions are close to the value of 3 which is frequently observed in mature stock markets.
Such values ensure a finite variance of returns.
This is important for option pricing and risk management.

\newpage

\begin{figure}[H]
\centering
\includegraphics[width=1.0\linewidth]{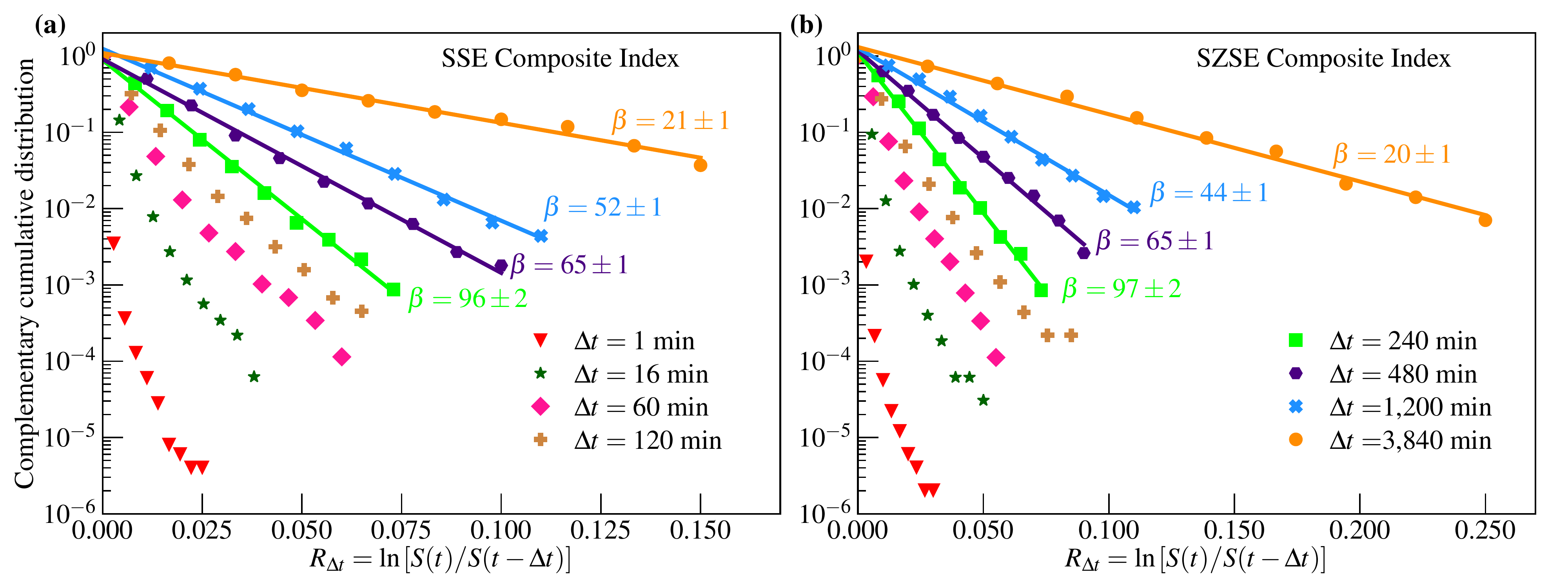}
\caption{\textbf{CCDF of return tails with different time scales $\Delta t$ in log--linear plot.}
Panels a and b reflect the SSECI and the SZSECI, respectively.
Colored markers are for positive tails.
Solid straight lines are exponential fits to data, and the values of $\beta$ with its fitting errors are also presented here.
These two panels share a common legend.
To keep the figure from looking cluttered, the negative tails are not shown here; they feature similar results to positive tails.}
\label{fig6}
\end{figure}

From Fig. \ref{fig5}, we find that the tail distributions with a large $\Delta t$ deviate from the asymptotic power-law.
However, we cannot obtain more details on the tails with a large $\Delta t$ since they are suppressed to the central region by normalization.
Here, we investigate the return tails in a log-linear style, as shown in Fig. \ref{fig6}.
Both positive and negative tails show exponential decay as the form of $e^{-\beta R_{\Delta t}}$ when $\Delta t \ge$ 240 min.
The exponential decay also ensures a finite variance of returns.
From Fig. \ref{fig5} and Fig. \ref{fig6},
we can conclude that the tails decay for the asymptotic power-law at a small value of $\Delta t$, and exponentially decay at large $\Delta t$ values,
which is not observed in mature stock markets.

\begin{figure}[H]
\centering
\includegraphics[width=1.0\linewidth]{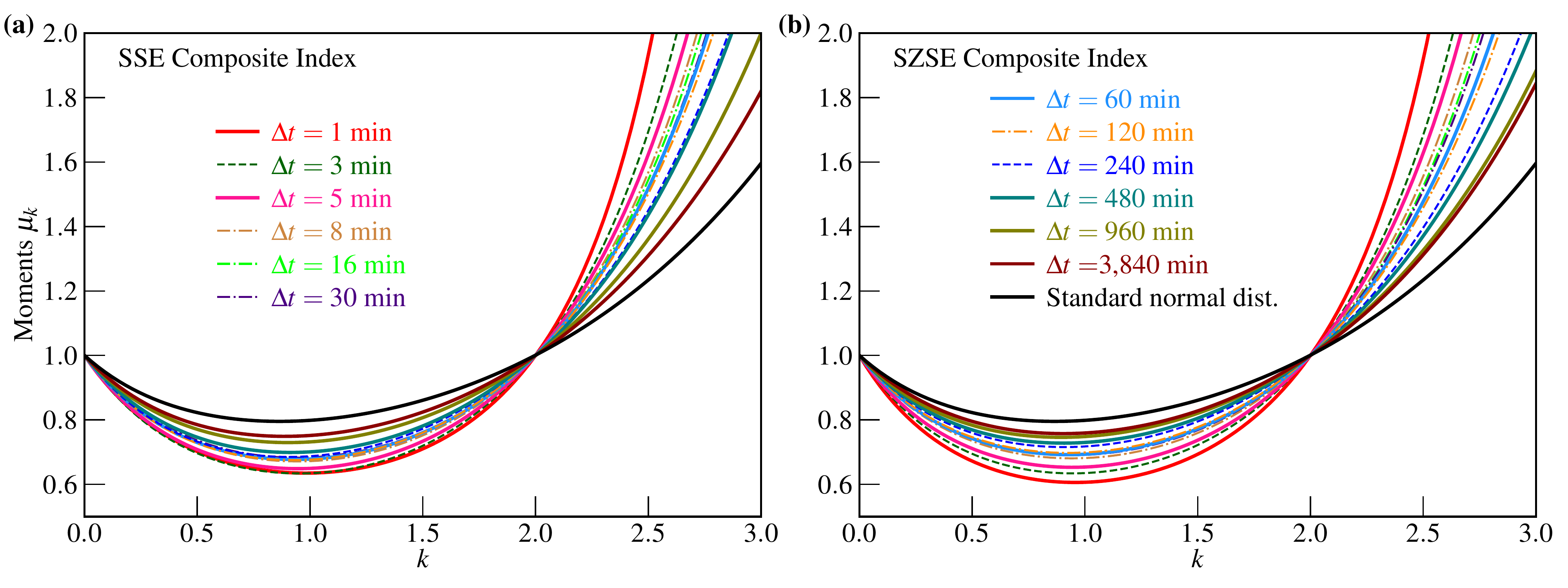}
\caption{\textbf{Comparison of moments $\mu_{k}$ between the normalized return data and the standard normal distribution for the SSECI (panel a) and the SZSECI (panel b).}
The colored curves denote the data, and the solid black curves refer to the moments of standard normal distribution. These two panels share a common legend.
It is evident that the data gradually converge to the standard normal distribution as $\Delta t$ increases.}
\label{fig7}
\end{figure}

\begin{figure}[H]
\centering
\includegraphics[width=1.0\linewidth]{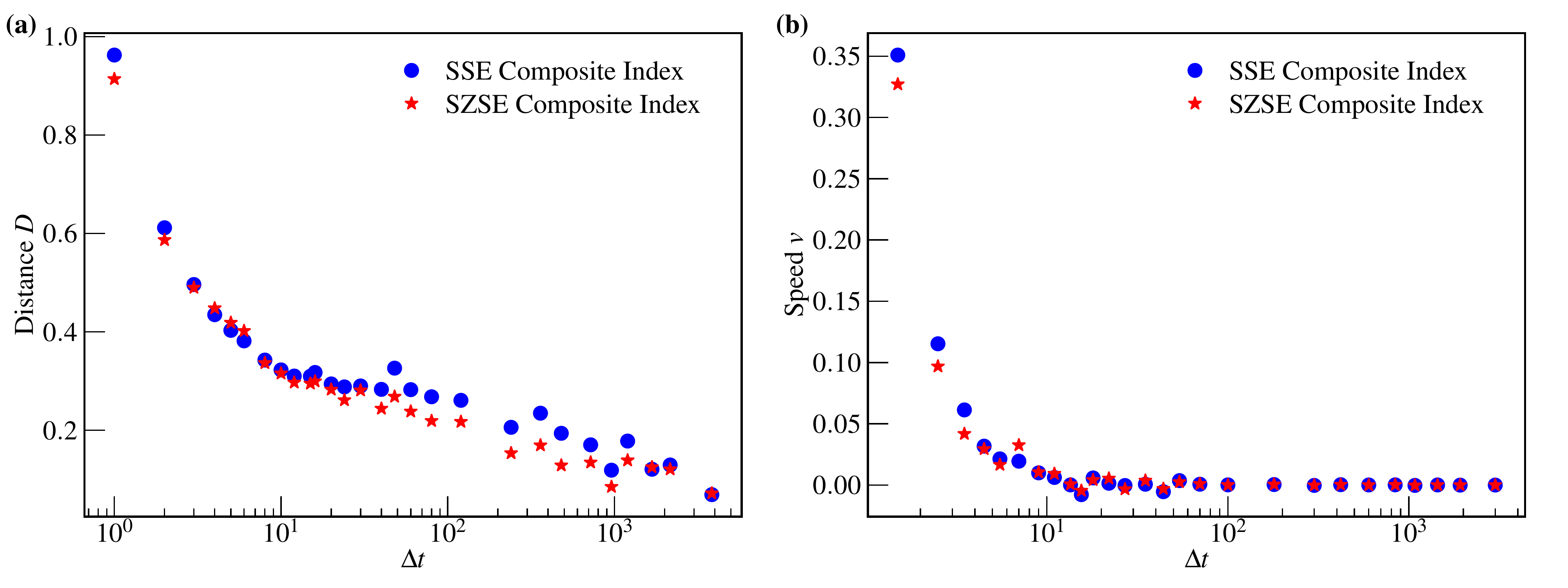}
\caption{\textbf{The moment distance between the normalized return data and the standard normal distribution (panel a) and the speed of convergence of data (panel b).}
$\Delta t$ is the time scale divided by 1 min. The circle and star points represent the data of SSECI and SZSECI, respectively.}
\label{fig8}
\end{figure}

We also verified the convergence behavior of return distribution by comparing the moments $\mu_{k}$ between the normalized return data and the standard normal distribution, as shown in Fig. \ref{fig7}.
The result indicates that the data gradually converge to the standard normal distribution starting from $\Delta t =$ 1 min.
To quantify this convergence behavior, we introduce a measure of moment difference between the normalized return data and the standard normal distribution, as shown in Eq. (\ref{distance}).
 \begin{equation}
 D = \sqrt{\frac{1}{n}\sum_{i=1}^{n}{\left[M_{D}\left(i\right) - M_G\left(i\right)\right]^{2}}}
 \label{distance}
 \end{equation}
where $M_{D}$ and $M_{G}$ denote the moments of the normalized return data with a $\Delta t$ and the standard normal distribution, respectively; $n$ is the number of data points shown in Fig. \ref{fig7}.
The measure of $D$ can also serve as the distance between two curves.
Therefore, we define speed using Eq. {\ref{speed}} to measure the speed of this convergence between a moment curve $i$ and another moment curve $i+1$, as shown in Fig. \ref{fig7}.
\begin{equation}
v = \frac{D_{i+1} - D_{i}}{\Delta t_{i+1} - \Delta t_{i}}
\label{speed}
\end{equation}
Fig. \ref{fig8} shows the measured distance between the normalized return data and the standard normal distribution and the speed of the convergence of data.
This figure demonstrates quantitatively that the convergence starts at $\Delta t =$ 1 min, and the speed at a small $\Delta t$ is much faster than others.
This convergence behavior is different from the early studies in the U.S.{\cite{PRE-1999-1}} and the Chinese stock markets{\cite{China-2008}},
in which convergence to the standard normal distribution occurs only when $\Delta t \ge$ 4 days{\cite{PRE-1999-1, China-2008}}.

\newpage

\section{Conclusions}
Because the previous studies on the return distribution of the Chinese stock market are dramatically limited by statistics,
this paper systematically and precisely investigates the property of the return distributions of both the SSECI and the SZSECI in the Chinese stock market.
We used 1 min high statistics datasets over a 17-year period (January 4, 2005 to December 31, 2021)
to construct return distributions with time intervals $\Delta t$ ranging from 1 min to almost 4,000 min.
The results illustrate that the properties of the return distributions for both the SSECI and the SZSECI are similar.
The main findings are as follows:
(1) The return distributions present a leptokurtic, fat-tailed, and almost symmetrical shape that is similar to that of mature stock markets.
(2) The central parts of the return distributions can be described by the symmetrical L\'{e}vy $\alpha$-stable process.
The key parameters $\alpha$ characterizing this process are extracted from our data.
They are 1.34 $\pm$ 0.03 (SSECI) and 1.13 $\pm$ 0.04 (SZSECI) over 1 $\le \Delta t \le$ 15 min
and 1.49 $\pm$ 0.03 (SSECI) and 1.57 $\pm$ 0.02 (SZSECI) over 60 $\le \Delta t \le$ 3,840 min.
Such values are comparable with the value of $\alpha \approx 1.4$ extracted from the U.S. stock market{\cite{Nature-1995}} and within the L\'{e}vy $\alpha$-stable process range of $0 < \alpha \le 2$.
(3) Return distributions can be described well by Student's t-distribution within a wider return range than the L\'{e}vy $\alpha$-stable distribution.
(4) A potential crossover region at 15 $< \Delta t <$ 60 min was discovered.
Such a crossover region is not observed in the U.S. stock market, where a single value of $\alpha \approx 1.4$ holds over 1 $\le \Delta t \le$ 1,000 min{\cite{Nature-1995}}. 
(5) To obtain a better understanding of tail distribution,
this paper checks the PDF and CCDF of tails in detail.
For small $\Delta t$, the tail shows scaling behavior and follows an asymptotic power-law decay with an exponent of about 3, which is a value widely observed in mature stock markets.
However, the tail decays exponentially when $\Delta t \ge$ 240 min, which is not observed in mature stock markets.
(6) Finally, it is observed that return distributions gradually converge to a normal distribution as $\Delta t$ increases.
Such convergence behavior is different from previous studies in the U.S {\cite{PRE-1999-1}}
and Chinese stock markets{\cite{China-2008}},
which state that convergence only occurs  when $\Delta t \ge$ 4 days.

Stock markets are inhomogeneous and time-varying.
A multifractal analysis via the return distribution and an analysis of volatility surfaces in stock markets across the world should be conducted using the latest high-frequency datasets that have been collected over the same time period.
By comparing the empirical results from different stock markets and constructing theoretical models, one can learn the underlying dynamics of stock markets{\cite{PNAS-2022}},
such as the impacts of investor risk attitude, trading rules, and government regulations in different countries.

\section*{Acknowledgements}
This work was supported by the Humanities and Social Sciences Youth Foundation of the Chinese Ministry of Education (Contract No. 22YJCZH107)
and the Scientific Research Support Program of Xi'an University of Finance and Economics (Contract No. 22FCZD03).
The authors would like to thank the anonymous referees and editors for their helpful comments which improve the quality of this paper.

\section*{Author contributions}
All authors made important contributions to this publication in the acquisition of data and data analysis.
Peng Liu wrote the manuscript.
All authors reviewed and approved the submitted manuscript.

\section*{Competing interests}
The authors declare no competing interests.

\end{document}